\documentstyle[icmp]{article}

\input{psfig}

\def\Journal#1#2#3#4{{#1} {\bf #2}, #3 (#4)}

\def\NPB{{\em Nucl. Phys.} B}
\def\PLB{{\em Phys. Lett.}  B}
\def\PRL{\em Phys. Rev. Lett.}
\def\PRD{{\em Phys. Rev.} D}

\def\be{\begin{equation}}
\def\ee{\end{equation}}
\def\bea{\begin{eqnarray}}
\def\eea{\end{eqnarray}}

\begin{document}

\title{DERIVATIVE EXPANSION OF THE ONE-LOOP EFFECTIVE ACTION IN QED}
 
\author{Igor~A.~Shovkovy}

\address{Physics Department, University of Cincinnati\\
         Cincinnati, OH 45221-0011, USA\\
         E-mail: igor@physics.uc.edu}

\maketitle\abstracts{The one-loop effective action in QED at zero and 
finite temperature is obtained by using the worldline approach. The 
Feynman rules for the perturbative expansion of the action in the number 
of derivatives are derived. The general structure of the temperature 
dependent part of the effective action in an arbitrary external 
inhomogeneous magnetic field is established. The two-derivative term 
in the effective action for spinor and scalar QED in a static magnetic 
background at $T\neq 0$ is calculated.}

The problem of calculating the effective action in QED is an old one. 
Its history starts with the well known papers by Heisenberg and Euler 
\cite{Heis}, and Weisskopf \cite{Wie}. Later, some results were 
obtained by Schwinger \cite{Sch} who, by using the proper time technique,
rederived the one-loop effective action for the case of a constant
electromagnetic field. Perhaps, the next most natural step in
solving the general problem is to take into account the effect of 
small deviations from a constant configuration of the field. 
It turns out, however, that the latter is very difficult to realize 
\cite{Hauk,Martin,Lee,DHok}. 

Here we present our recent work \cite{GS} that generalizes the previously 
known results on the effective action in QED. There, in particular, 
the derivative expansions up to two derivatives of the field 
strength with respect to space-time coordinates in scalar and spinor 
QED at zero and at finite temperature are obtained (for some partial 
results in a non-Abelian gauge theory see \cite{McK,GarMcA}). 

Our method is heavily based on the very elegant and now widely 
developed worldline approach to quantum field theory 
\cite{BernK,Stras,McK1,McK2,Schmidt,vanH,RSS,FHSS}. Here we
follow a self-contained approach of Ref.~\cite{GS}, so that 
even a non-expert in the field should understand all the 
details. 

As is known, the one-loop effective action in QED reduces to 
computing the fermion determinant
\begin{eqnarray}
W^{(1)}(A)&=&-i\ln {\rm Det}(i\hat{{\cal D}}-m)=
-\frac{i}{2}\ln{\rm Det}\left({\cal D}^2_{\mu}
+\frac{e}{2}\sigma_{\mu\nu}F^{\mu\nu}+m^2\right)
\label{eq:ActGen2}\\
&=&-\frac{i}{2}\int d^{n}x\langle x|tr\ln\left({\cal D}^2_{\mu}+
\frac{e}{2}\sigma_{\mu\nu}F^{\mu\nu}+m^2\right) |x\rangle
\equiv  \int d^{n}x{\cal L}^{(1)},\nonumber
\end{eqnarray}
where
\begin{equation}
{\cal L}^{(1)}(A)=\frac{i}{2} \int \limits^{\infty }_{0} 
\frac{d\tau}{\tau} e^{-im^2\tau} tr \langle x| \exp(-i\tau H) 
|x\rangle.
\label{eq:ActGen3}
\end{equation}
Here $\hat{{\cal D}}=\gamma^{\mu}{\cal D}_{\mu}$ and the covariant
derivative is ${\cal D}_{\mu}= \partial_{\mu}+ieA_{\mu}$. By
definition, $\sigma_{\mu\nu}=i[\gamma_{\mu},\gamma_{\nu}]/2$ and 
$tr$ refers to the spinor indices of the Dirac matrices
$\gamma_{\mu}$. States $|x\rangle$ are the eigenstates of a
self-conjugate coordinate operator $x_{\mu}$. Throughout the paper
we use the Minkowski metric, i.e., $\eta_{\mu\nu}=(1,-1,-1)$ or
$\eta_{\mu\nu}=(1,-1,-1,-1)$, depending on the actual space-time
dimension. And in both $2+1$ and $3+1$ dimensions, we work with 
the $4\times 4$ representation of the Dirac $\gamma$-matrices. 

At finite temperature, the expression analogous to that in 
Eqs.~(\ref{eq:ActGen2}) reads
\begin{eqnarray}
F^{(1)}(A)&=&-T \ln {\rm Det}(i\gamma^{\mu}{\cal D}_{\mu}-m)=
-\frac{T}{2}\ln{\rm Det}\left({\cal D}^{\mu}{\cal D}_{\mu}
+\frac{e}{2}\sigma^{\mu\nu}F_{\mu\nu}+m^2\right)=\nonumber\\
&=&-\frac{T}{2}\int\limits_{0}^{-i\beta} dx_{0}
\int d^{3}x\langle x|tr\ln\left({\cal D}^{\mu}{\cal D}_{\mu}
+\frac{e}{2}\sigma^{\mu\nu}F_{\mu\nu}+m^2\right) |x\rangle,
\label{ActGen2}
\end{eqnarray}
so that
\begin{equation}
\frac{F^{(1)}(A)}{V_3}=-\frac{i}{2} \int \limits^{\infty }_{0} 
\frac{d\tau}{\tau} e^{-im^2\tau} tr \langle x| \exp(-i\tau H) 
|x\rangle.
\label{free-e}
\end{equation}
In both cases, $T=0$ and $T\neq 0$, the second order differential 
operator $H$ is given by
\begin{equation}
H= {\cal D}^{\mu} {\cal D}_{\mu} + \frac{e}{2} \sigma^{\mu\nu} 
F_{\mu\nu}(x).
\label{operH}
\end{equation}
The matrix element $\langle z|U(\tau)|y\rangle
\equiv \langle z|\exp(-i\tau H)|y\rangle$ 
(which has the interpretation of the evolution operator of
a spinning particle), entering the right hand side of 
Eq.~(\ref{free-e}), allows a quantum mechanical path integral 
representation,
\begin{equation}
tr \langle z|U(\tau)|y\rangle = \frac{1}{N}\int {\cal D} 
[x(t), \psi(t)]\exp\left\{ i\int \limits^{\tau}_{0} dt 
\left[ L_{bos}(x) + L_{fer}(\psi,x) \right] \right\},
\label{evol0}
\end{equation}
where $N$ is a normalization factor, and
\begin{equation}
L_{bos}(x)= -\frac{1}{4}\frac{dx_{\nu }}{dt}
\frac{dx^{\nu }}{dt}-eA_{\nu }(x) \frac{dx^{\nu }}{dt},
\label{L_bos}
\end{equation}
\begin{equation}
L_{fer}(\psi ,x)=\frac{i}{2}\psi_{\nu }\frac{d\psi^{\nu }}{dt}
-ie \psi^{\nu }\psi^{\lambda } F_{\nu \lambda }(x).
\label{L_fer}
\end{equation}
The integration in Eq.~(\ref{evol0}) goes over trajectories 
$x^{\mu}(t)$ and $\psi^\mu (t)$ parameterized by $t\in
[0,\tau]$. In addition, the definition of the integration 
measure assumes the following boundary conditions
\begin{eqnarray}
x_{\nu }(0)=y_{\nu },&\qquad & \psi (0)=-\psi (\tau) \\
x_{0}(\tau)=z_{0}~\mbox{mod}(i\beta),&\qquad &
x_{i}(\tau)=z_{i} \quad (i=1,2,3).
\end{eqnarray}
Note that, at zero temperature, $\mbox{mod}(i\beta)$ is absent in the
boundary conditions.

Since the path integral (in the case of finite temperature) 
includes the integrations over 
worldline trajectories with arbitrary integer windings
around the compact (imaginary) $x_0$-direction, the expression 
in Eq.~(\ref{evol0}) splits into the sum of path integrals 
labeled by the winding numbers. In case of spinor QED, the weight 
factors of these separate contributions are given \cite{McK1}
by $(-1)^n$. Therefore,
\begin{eqnarray}
&& tr\langle z|U(\tau) |y\rangle =
\frac{1}{N}\sum_{n=-\infty}^{\infty}
(-1)^n \int {\cal D} [x^{(n)}(t) , \psi(t) ] 
\nonumber\\
&\times&\exp \left \{i\int \limits^{\tau}_{0} dt 
\left[L_{bos} \left(x^{(n)}(t) \right)
+L_{fer} \left( \psi(t), x^{(n)}(t) \right) \right] \right\},
\label{evol1}
\end{eqnarray}
where the boundary conditions $x^{(n)}_{\nu }(0)=y_{\nu }$ and 
$x^{(n)}_{\nu }(\tau)=z_{\nu }+in\beta\eta_{\nu 0}$ are assumed. 
Note, that there exists a similar representation for scalar QED 
as well. In contrast to the case at hand, in scalar QED, the 
integration over the Grassman field $\psi(t)$ is absent and all 
the weight factors are equal \cite{McK1} to 1. At zero temperature 
we have only $n=0$ term, and the sum does not appear at all. We 
will see later, Eqs.~(\ref{inB3+1}) and (\ref{csB3+1}), that the 
same is true if we consider the limit $T\to 0$ ($\beta\to\infty$) 
afterwards, i.e., all the terms with $n\neq 0$ go to zero.

Let us consider the effective action in the case of a slightly 
inhomogeneous static magnetic field. We choose a version of the 
Fock-Schwinger \cite{GS} gauge for the vector potential $A_{\mu}(x)$, 
\begin{equation}
A_{0}(x)=0, \qquad (x_{i}-y_{i})A_{i}(x)=0.
\end{equation}
The latter leads to the series
\begin{eqnarray}
A_{i}(x) &=& -\frac{1}{2} (x_{j} -y_{j}) 
F_{ji}(y) + \frac{1}{3} (x_{j}-y_{j})
(x_{l}-y_{l}) \partial _{l} F_{ji}(y)
\nonumber\\ &-& 
\frac{1}{8} (x_{j}-y_{j}) (x_{l}-y_{l}) 
(x_{k}-y_{k}) \partial _{l}\partial _{k} 
F_{ji}(y) + \ldots.
\label{eq:Ai}
\end{eqnarray}
With this choice of gauge, we arrive at a very convenient 
representation for the diagonal matrix element of the evolution 
operator, 
\begin{eqnarray} 
&&tr\langle y|U(\tau)|y \rangle= \sum_{n=-\infty}^{\infty}
\frac{(-1)^n}{N} \int {\cal D} [x^{(n)}, \psi] 
\exp\left[i\int\limits^\tau_0 dt \left(
-\frac{1}{4}\frac{d x^{(n)}_{0}}{dt}\frac{d x^{(n)}_{0}}{dt}
\right)\right]
\nonumber\\ 
&\times& \exp\left[i\int\limits^\tau_0 dt \left(
\frac{1}{4}\frac{d x^{(n)}_{i}}{dt}\frac{d x^{(n)}_{i}}{dt} 
-\frac{e}{2}x^{(n)}_{i} F_{ij}(y) \frac{d x^{(n)}_{j}}{dt}
+L_{bos}^{int}\left(x^{(n)}_{i}\right)\right)\right]
\label{trU1}\\ 
&\times&\exp\left[i\int\limits^{\tau}_{0}dt \left(
\frac{i}{2}\psi_{0} \frac{d\psi_{0}}{dt}
-\frac{i}{2}\psi_{i} \frac{d\psi_{i}}{dt}
-ie\psi_{i} \psi_{j} F_{ij}(y) 
+L_{fer}^{int}\left(x^{(n)}_{i},\psi_{i}\right) 
\right)\right],
\nonumber
\end{eqnarray}
where, as follows from Eqs.~(\ref{L_bos}), (\ref{L_fer}) and 
(\ref{eq:Ai}), the interacting terms, $L_{bos}^{int}(x)$ and 
$L_{fer}^{int}(x,\psi)$, containing spatial derivatives of 
$F_{ij}$, are given by
\begin{eqnarray}
L_{bos}^{int}(x)&=& 
- \frac{e}{3} F_{i j,k}\frac{d x_{i}}{dt} x_{j} x_{k}
+ \frac{e}{8} F_{i j,k l} \frac{d x_{i}}{dt} x_{j} x_{k} x_{l} 
+ \ldots,   
\label{eq:L2}\\
L_{fer}^{int}(x,\psi ) &=&
i eF_{i j,k} \psi_{i} \psi_{j} x_{k}
- \frac{ie}{2} F_{i j,k l} \psi_{i} \psi_{j} x_{k} x_{l} 
+ \ldots.
\label{eq:L3}
\end{eqnarray}
The integration variables in Eq.~(\ref{trU1}) are subject 
to the following boundary conditions, $x^{(n)}_{0,i}(0)=0$, 
$x^{(n)}_{0}(\tau)=in\beta$ and $x^{(n)}_{i}(\tau)=0$ (note 
that the fields $x^{(n)}_{0,i}(t)$ were preliminary shifted by 
$-y_{0,i}$).

A very nice property of the path integral in Eq.~(\ref{trU1}) 
is its factorization into two pieces. One of them contains only 
the time components of the fields and is, in fact, a Gaussian path 
integral. The other contains the interacting spatial components 
of the fields and, what is very important, does not depend on the 
winding number $n$. After performing the Gaussian integrations 
over $x^{(n)}_{0}(t)$ and $\psi_{0}(t)$, we obtain
\begin{eqnarray} 
&&tr\langle y|U(\tau)|y \rangle=
\frac{1}{N}\sum_{n=-\infty}^{\infty}
(-1)^n\exp\left(i\frac{n^2\beta^2}{4\tau}\right)
\int {\cal D} [x_{i}(t), \psi_{i}(t) ] \nonumber\\ 
&\times& \exp\left[i\int\limits^\tau_0 dt \left(
\frac{1}{4}\frac{d x_{i}}{dt}\frac{d x_{i}}{dt} 
-\frac{e}{2}x_{i} F_{ij}(y) \frac{d x_{j}}{dt}
+L_{bos}^{int}\left(x_{i}\right)\right)\right]
\nonumber\\ 
&\times&\exp\left[i\int\limits^{\tau}_{0}dt \left(
-\frac{i}{2}\psi_{i} \frac{d\psi_{i}}{dt}
-ie\psi_{i} \psi_{j} F_{ij}(y) 
+L_{fer}^{int}\left(x_{i},\psi_{i}\right) 
\right)\right]. 
\label{factor}
\end{eqnarray}
This expression, in fact, is one of our most important results 
here. It states that the temperature dependent part of 
the evolution operator is exactly factorized from the part 
depending on an (arbitrary) inhomogeneous magnetic field. 
The latter, in its turn, puts a strong restriction on the 
structure of the one-loop finite temperature effective action 
in QED. Similarly, we obtaine the result in scalar QED,
\begin{eqnarray} 
&&tr\langle y|U_{scal}(\tau)|y \rangle=
\frac{1}{N}\sum_{n=-\infty}^{\infty}
\exp\left(i\frac{n^2\beta^2}{4\tau}\right)
\int {\cal D} [x_{i}(t), \psi_{i}(t) ] \nonumber\\ 
&\times& \exp\left[i\int\limits^\tau_0 dt \left(
\frac{1}{4}\frac{d x_{i}}{dt}\frac{d x_{i}}{dt} 
-\frac{e}{2}x_{i} F_{ij}(y) \frac{d x_{j}}{dt}
+L_{bos}^{int}\left(x_{i}\right)\right)\right]. 
\label{factor-bos}
\end{eqnarray}
As is seen, the path integrals in Eqs.~(\ref{factor}) and 
(\ref{factor-bos}) are almost the same as those at zero 
temperature. The only difference is due to the additional 
(Gaussian) integration over $x_0$ at $T=0$. The latter, 
however, is irrelevant since it just modifies the overall 
normalization factor $1/N$. 

The further integration (over the spatial components of the 
fields) can be done only approximately. Remarkably, the result 
in the case of a constant background field can be obtained exactly 
from the above expressions \cite{GS}. That is due to the fact that 
the integrals in Eqs.~(\ref{factor}) and (\ref{factor-bos}) 
are Gaussian in that particular case. All the derivatives 
of an inhomogeneous background field appear in the worldline 
action only through the interaction terms. Therefore, it 
is convenient to develop the corresponding perturbative theory
and formulate the Feynman rules for calculating the diagrams of 
interest. Here we briefly outline these rules.

\begin{figure}[t]
\psfig{figure=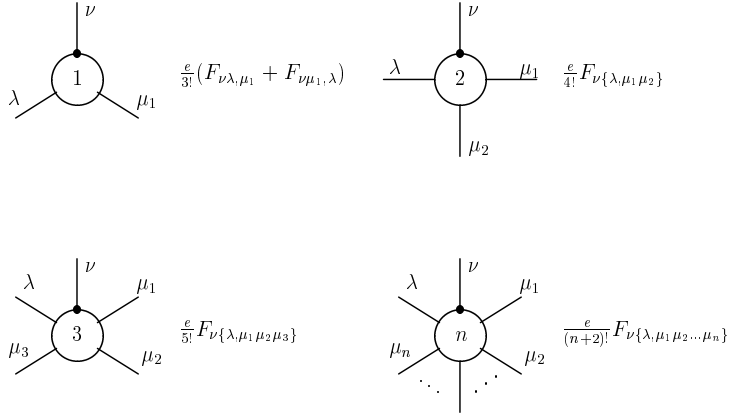,height=2.5in}
\caption{Diagrammatic notations for the boson interaction vertices.  
The curly brackets denote symmetrization of the type:  
$F_{\nu\{\lambda,\mu_1\dots\mu_n\}} 
=F_{\nu\lambda,\mu_1\dots\mu_n} 
+F_{\nu\mu_1,\lambda\dots\mu_n} +\dots 
+F_{\nu\mu_n,\mu_1\dots\lambda}$. \label{Figure1}}
\end{figure}

\begin{figure}[t]
\psfig{figure=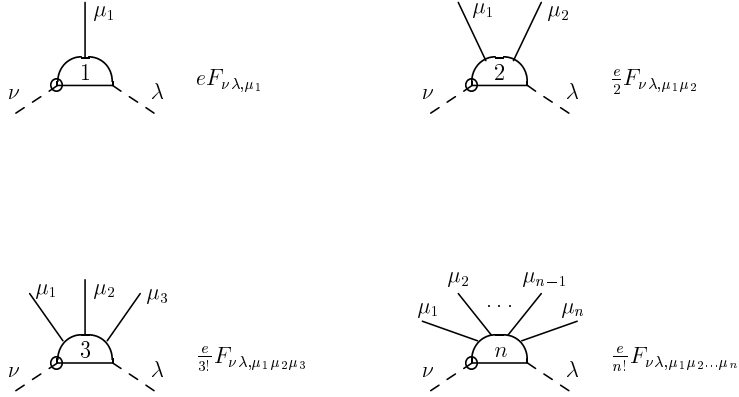,height=2.5in}
\caption{Diagrammatic notations for the fermion-boson 
interaction vertices. \label{Figure2}}
\end{figure}

We observe that there are two different types of local interactions 
in Eqs.~(\ref{eq:L2}) and (\ref{eq:L3}). The first (bosonic) type
contains only the bosonic fields, $x_{\mu}(t)$. The corresponding 
vertices are shown in Figure~\ref{Figure1}. The other type involves 
both the boson, $x_{\mu}(t)$, and the spinor fields, $\psi_{\mu}(t)$.  
These latter produce the vertices given in Figure~\ref{Figure2}. 
Regarding the notation, the integers in the vertices denote the 
number of derivatives of the electromagnetic field with respect to 
space-time coordinates. Some legs in the diagrams are marked by 
circles and bullets. The circles correspond to legs related to 
the first Lorentz index ($\nu$) of the tensor weight, 
$F_{\nu\lambda,\mu_1,\dots,\mu_n}$, assigned to the fermion 
vertices. The bullets, on the other hand, mark those legs 
which contain the derivatives with respect to the proper time.  
The latter act on the boson propagators attached to the
corresponding legs. 

The rest of the Feynman rules are easy to derive as well~\cite{GS}. 
In calculation, one has to use the appropriate propagators 
of $x_{\mu}(t)$ and $\psi_{\mu}(t)$ fields \cite{GS} for connecting 
the boson (solid) and the fermion (dashed) legs, respectively.  

A somewhat disappointing feature of the theory is an infinite number 
of local interactions (see Eqs.~(\ref{eq:L2}) and (\ref{eq:L3})). 
Fortunately, while working at any finite order of the perturbative 
theory, one requires only a finite number of those interactions. 

Omitting the intermediate calculations, let us give the final 
results for the first non-trivial term in the derivative 
expansion  of the effective action (free energy) in QED.
Assuming that the background magnetic field is directed
along the third spatial axis, we get the following expression 
for the two-derivative part of the expansion,
\begin{eqnarray} 
\frac{F^{(1)}_{der}(B)}{V_3}&=&
\frac{e^2 \left(\partial_{\perp}B\right)^2}{(8\pi)^{2}|eB|} 
\int\limits_{0}^{\infty} \frac{d\omega}{\omega} 
e^{-m^2\omega/|eB|} 
\frac{d^3}{d\omega^3}\left( \omega \coth\omega \right)
\nonumber\\
&\times& \sum_{n=-\infty}^{\infty} (-1)^n
\exp\left( -\frac{n^2\beta^2|eB|}{4\omega}
\right),
\label{inB3+1}
\end{eqnarray} 
where $\left(\partial_{\perp}B\right)^2\equiv
\left(\partial_{1}B\right)^2+\left(\partial_{2}B\right)^2$.
The result in Eq.~(\ref{inB3+1}) differs from the analogous 
expression at $T=0$ (see Ref.~\cite{GS}) only by the last 
factor containing the sum over the winding numbers. 
Again, as all the other results, this expression allows 
a straightforward generalization to 2+1 dimensional case as well 
as to scalar QED. For example, in 3+1 dimensional scalar QED,
the result reads 
\begin{eqnarray} 
\frac{F^{(1)scal}_{der}(B)}{V_3}&=&
-\frac{e^2 \left(\partial_{\perp}B\right)^2}{2(8\pi)^{2}|eB|} 
\int\limits_{0}^{\infty} \frac{d\omega}{\omega} 
e^{-m^2\omega/|eB|} 
\left(\frac{d^3}{d\omega^3}+\frac{d}{d\omega} \right)
\left( \frac{\omega }{\sinh\omega }\right)\nonumber\\
&\times&\sum_{n=-\infty}^{\infty}
\exp\left( -\frac{n^2\beta^2|eB|}{4\omega}
\right). 
\label{csB3+1}
\end{eqnarray}

In conclusion, here we derived the one-loop zero and finite 
temperature effective potential (free energy) in spinor and scalar 
QED  using the worldline formulation of quantum field theory. 
The only difference (at one loop) between zero and finite 
temperature cases appears in applying the boundary conditions to 
the saddle point trajectory. By making use of this method, we were 
able to established the general structure of the temperature dependent
part of the effective action in QED in an arbitrary external 
inhomogeneous magnetic field. Also, we  established the Feynman 
rules for calculating the perturbative expansion of the effective 
action (free energy) in the number of derivatives of the background 
field with respect to space-time coordinates. The explicit result 
for the first non-trivial term in the expansion (containing two 
derivatives of the magnetic field) is presented.

\section*{Acknowledgments}
The author would like to thank V.~Gusynin for helpful comments 
on the manuscript. This work was supported in part by the 
U.S. Department of Energy Grant \#DE-FG02-84ER40153.


\section*{References}
\begin{thebibliography}{99}

\bibitem{GS} V.P.~Gusynin and I.A.~Shovkovy,  
\Journal{\em Can. J. Phys.}{74}{282}{1996}; 
Derivative Expansion of the Effective Action for QED 
in 2+1 and 3+1 dimensions, {\tt hep-th/9804143}; 
I.A.~Shovkovy, \Journal{\PLB}{441}{313}{1998}, {\tt hep-th/9806156}.

\bibitem{Heis} W.~Heisenberg, H.~Euler,  
\Journal{\em Z. Phys.}{98}{714}{1936}.

\bibitem{Wie} V.~Weisskopf, 
\Journal{\em K. Dan. Vidensk. Selsk. Mat.-Fys. Medd.}{14}{6}{1936}.

\bibitem{Sch} J.~Schwinger, \Journal{\em Phys. Rev.}{82}{664}{1951}.

\bibitem{Hauk} J.~Hauknes, \Journal{\em Ann. Phys.}{156}{303}{1984}.

\bibitem{Martin} C.~Martin and D.~Vautherin, 
\Journal{\PRD}{38}{3593}{1988}.

\bibitem{Lee} H.W.~Lee, P.Y.~Pac, and H.K.~Shin,  
\Journal{\PRD}{40}{4202}{1989}.

\bibitem{DHok} D.~Cangemi, E.~D'Hoker, and G.~Dunne, 
\Journal{\PRD}{51}{R2513}{1995}.

\bibitem{McK} D.G.C.~McKeon, 
\Journal{\PRD}{55}{7989}{1997}.

\bibitem{GarMcA} T.D.~Gargett and I.N.~McArthur, 
\Journal{\em J. Math. Phys.}{39}{4430}{1998}.

\bibitem{BernK} Z.~Bern and D.A.~Kosower, 
\Journal{\PRL}{66}{1669}{1991};
\Journal{\NPB}{379}{451}{1992}.  

\bibitem{Stras} M.J.~Strassler, 
\Journal{\NPB}{385}{145}{1992}. 

\bibitem{McK1} D.G.C.~McKeon and A.~Rebhan,
\Journal{\PRD}{47}{5487}{1993};
\Journal{\PRD}{49}{1047}{1994}.

\bibitem{McK2} D.G.C.~McKeon,  
\Journal{\em Can. J. Phys.}{71}{334}{1993};
\Journal{\em Ann. Phys.}{224}{139}{1993}; 
\Journal{{\em Int. J. Mod. Phys.} A}{12}{5387}{1997}. 

\bibitem{Schmidt} M.G.~Schmidt and C.~Schubert,
\Journal{\PLB}{318}{438}{1993}; 
\Journal{\PLB}{331}{69}{1994}; 
\Journal{\PRD}{53}{2150}{1996}.

\bibitem{vanH} J.W.~van~Holten, 
\Journal{\NPB}{457}{375}{1995}.  

\bibitem{RSS} M.~Reuter, M.G.~Schmidt and C.~Schubert,
\Journal{\em Ann. Phys.}{259}{313}{1997}.   

\bibitem{FHSS} D.~Fliegner, P.~Haberl, M. G.~Schmidt 
and C.~Schubert, \Journal{\em Ann. Phys.}{264}{51}{1998}.

\end{thebibliography}
\end{document}